\documentstyle[prl,aps,multicol,epsf,rotate]{revtex}
\begin{document}
\draft
\preprint{XXXX}
\title{Stretched exponential relaxation in the
mode-coupling theory for the Kardar-Parisi-Zhang equation}
\author{Francesca Colaiori and M. A. Moore}
\address{Department of Physics and Astronomy, University of 
Manchester,
Manchester, M13 9PL, United Kingdom}
\date{\today}
\maketitle
\begin{abstract}
We study the mode-coupling theory for the Kardar-Parisi-Zhang
equation in the strong-coupling regime, focusing on the long
time properties. By a saddle point analysis of the mode-coupling
equations, we derive exact results for the correlation function
in the long time limit -- a limit  which is hard to study using 
simulations. The correlation function at wavevector ${\bf k}$ in 
dimension $d$ is found to behave asymptotically at time $t$ as
$C({\bf {k}},t)\simeq \frac{A}{k^{d+4-2z}} (Btk^z)^{\gamma/z}
e^{-(Btk^z)^{1/z}}$, with $\gamma=(d-1)/2$, $A$ a determined constant 
 and $B$ a scale factor.
\end{abstract}
\begin{multicols}{2}
%
%\cite{}
The Kardar-Parisi-Zhang (KPZ) \cite{KPZ} equation is a simple
non-linear Langevin equation, proposed in 1986 as a coarse grained
description of a growing interface. Probably due to the fact
that is the simplest generalization
of the diffusion equation which includes a relevant nonlinear 
term,
the KPZ equation also arises in  connection with many other
important physical problems (the Burgers equation for 
one-dimensional turbulence \cite{For},
directed polymers in a random medium \cite{Kar,Der,Par} etc.). 

The KPZ equation for a growing interface, described by a single
valued height function $h({\bf x},t)$ on a $d$-dimensional 
substrate
${\bf x}\in \Re^d$ is:
\begin{equation}
\partial_t h({\bf x},t)=
\nu \nabla^2 h +\frac{\lambda}{2}(\nabla h)^2+\eta({\bf x},t) \,.
\label{KPZ}
\end{equation}
The first term represents the surface tension forces which tend to
smooth the interface, the second describes the nonlinear growth 
locally
normal to the surface, and the last is a noise which mimics the
stochastic nature of the growth process \cite{rev}.
We choose the noise to be Gaussian, with zero mean and second 
moment
$\langle \eta({\bf x},t)\eta({\bf x'},t')\rangle=
2D\delta^{d}({\bf x}-{\bf x'})\delta(t-t') .$
The steady state interface profile is usually described in terms
of the roughness:
$w=\langle h^2({\bf x},t)\rangle- \langle h({\bf x},t)\rangle^2 $
which for a system of size $L$ behaves like $L^{\chi} f(t/L^z)$,
where $f(x)\rightarrow const$ as $x\rightarrow \infty$
and $f(x) \sim x^{\chi/z}$ as $x \rightarrow 0$, so that $w$
grows with time like $t^{\chi/z}$ until it saturate to $L^{\chi}$
when $t\sim L^z$. $\chi$ and $z$ are the roughness and dynamic
exponent respectively. 

The phenomenology of the KPZ is well known:
above two  dimensions, there are two distinct regimes,
separated by a critical value $\lambda_c$ of the non-linearity
coefficient.
In the weak coupling regime ($\lambda < \lambda_c$) the non-linear
term is irrelevant and the behavior
is governed by the Gaussian ($\lambda=0$) fixed point. The KPZ
in this regime is equivalent to the
linear Edward Wilkinson equation, for which the
exponents are known exactly $\chi=(2-d)/2$ and $z=2$.
The more challenging strong coupling regime ($\lambda > 
\lambda_c$),
where the non-linearity is relevant is characterised by anomalous
exponents, which are not known exactly in general dimension $d$.
From the Galilean invariance \cite{For}
(invariance of Eq. (\ref{KPZ}) under an infinitesimal
tilting of the surface) one can derive the relation
$\chi+z=2$, which leaves just one independent exponent. For the
special case when $d=1$, the existence of a 
fluctuation-dissipation
theorem gives the exact results 
$\chi=1/2$, $z=3/2$.

We take a self-consistent approach, the mode-coupling (MC) 
approximation
\cite{For}, in which in the diagrammatic expansion for the 
correlation
and response function only diagrams which do not renormalise
the three point vertex $\lambda$ are retained. The MC 
approximation
has been remarkably successful in other
areas of condensed matter physics, for example in the study of 
structural
glass transitions \cite{gl},  binary mixtures \cite{bm} and 
critical
dynamics of magnets \cite{bm,cdm}.
KPZ mode-coupling reproduces the exact values of the exponents
in $d=1$. Furthermore MC equations have been shown
to arise from the large $N$-limit of an generalised $N$-component 
KPZ equation
\cite{Do}, which allows in principle a systematic approach to the
theory beyond mode coupling, by expanding in $1/N$.

Our study focuses on the large time properties of the KPZ 
equation:
we predict a stretched exponential decay for the correlation
function at large times.
A phenomenological stretched exponential law was used as long ago 
as 1854 to fit electronic relaxation data for a Leiden jar capacitor 
\cite{vis}, and then been rediscovered many times: fitting functions 
involving stretched exponentials are nowadays widely used in phenomenogical
analysis of relaxation data (examples are dielectric relaxation
\cite{die} and  glassy relaxation \cite{gl,gla}).
However, only a few analytical arguments \cite{arg} are able to
reproduce stretched exponential relaxation in complex systems.
Our prediction is in principle amenable of numerical verification,
even though usual numerical techniques, mainly based on simulations
\cite{Num} of discrete microscopical models which belong to the KPZ
universality class \cite{Mar}, are hard in the long time
asymptotic region.

Mode-coupling equations are coupled equation for
the correlations and response function.
The correlation and response function are defined in
Fourier space by
\end{multicols}	
%%%%%%%%%%%%%%%%%%%%%%%%%%%%%%%%%%%%%%%%%%%%%%%%%%%%%%%%%%%%%%%%%%
%%%%%%%%%%%%%%%%%%%%%%%%%%%%%%%%%%%%%%%%%%%%%%%%%%%%%%%%%%%%%%%%%%
%%%%%%%%%%%%%%%%%%%%%%%%%%%%%%%%%%%%%%%%%%%%%%%%%%%%%%%%%%%%%%%%%%
%%%%%%%%%%%%%%%%%%%%%%%%%%%%%%%%%%%%%%%%%%%%%%%%%%%%%%%%%%%%%%%%%%
%%%%%%%%%%%%%%%%%%%%%%%%%%%%%%%%%%%%%%%%%%%%%%%%%%%%%%%%%%%%%%%%%%
%%%%%%%%%%%%%%%%%%%%%%%%%%%%%%%%%%%%%%%%%%%%%%%%%%%%%%%%%%%%%%%%%%
\begin{eqnarray}
&C({\bf k},\omega)=\langle h({\bf k},\omega) h^{*}({\bf k},\omega) 
\rangle,
\nonumber
\\
&G({\bf k},\omega)=\delta^{-d}({\bf k}-{\bf k}^{'})\delta^{-1}
(\omega-\omega^{'})
\langle
\frac{\partial h({\bf k},\omega)}{\partial \eta({\bf 
k}^{'},\omega^{'})}
\rangle ,
\nonumber
\end{eqnarray}
\noindent
where $\langle \cdot \rangle$ indicate an average over $\eta$.
In the mode-coupling approximation, the correlation and response
functions are the solutions of two coupled equations,
%\end{multicols}
\begin{eqnarray}
&
G^{-1}({\bf k},\omega)=G^{-1}_0({\bf k},\omega)+\lambda^2
\int \frac{d\Omega}{2 \pi} \int \frac{d^dq}{(2 \pi)^d}
\left[{\bf q} \cdot ({\bf k}-{\bf q})\right]\left[{\bf q} \cdot 
{\bf
k}\right]
G({\bf k}-{\bf q},\omega - \Omega) C({\bf q},\Omega) \,,
\label{mc1}
\\&
C({\bf k},\omega)=C_0({\bf k},\omega)+\frac{\lambda^2}{2}
\mid G({\bf k},\omega)\mid^2
\int \frac{d\Omega}{2 \pi} \int \frac{d^dq}{(2 \pi)^d}
\left[{\bf q} \cdot ({\bf k}-{\bf q})\right]^2
C({\bf k}-{\bf q},\omega - \Omega) C({\bf q},\Omega) \,,
\label{mc2}
\end{eqnarray}
%\begin{multicols}{2}
\noindent
where $G_0({\bf k},\omega)=(\nu k^2 - i \omega)^{-1}$ is the bare
response function, and $C_0({\bf k},\omega)=2 D \mid G({\bf
k},\omega)\mid^2$.
In the strong coupling limit,
$G({\bf k},\omega)$ and $C({\bf k},\omega)$ take the following
scaling forms
\begin{eqnarray}
&
G({\bf k},\omega)=k^{-z}g\left( \omega/k^{z}\right) \,,
\nonumber
\\&
C({\bf k},\omega)=k^{-(2 \chi+d+z)}n\left( \omega/k^{z}\right) \,,
\nonumber
\end{eqnarray}
\noindent
and Eqs. (\ref{mc1}) and  (\ref{mc2})
translate into the following coupled
equations for the scaling functions $n(x)$ and $g(x)$:
\begin{eqnarray}
&
g^{-1}(x)=-i x +I_{1}(x) \,,
\label{g}
\\
&
n(x)=\mid g(x)\mid^{2}I_{2}(x) \,,
\label{n}
\end{eqnarray}
where $x=\omega/k^z$ and $I_1(x)$ and $I_2(x)$ are given by
%\end{multicols}
\begin{eqnarray}
&
I_1(x)=P\int_{0}^{\pi}d\theta \sin^{d-2}\theta
\int_{0}^{\infty}dq\cos\theta (\cos\theta-q)q^{2z-3}r^{-z}
\int_{-\infty}^{\infty}dy 
\,g\left(\frac{x-q^{z}y}{r^{z}}\right)n(y),
\nonumber
\\
&
I_2(x)=\frac{P}{2}\int_{0}^{\pi}d\theta \sin^{d-2}\theta
\int_{0}^{\infty}dq(\cos\theta-q)^2 q^{2z-3}r^{-(d+4-z)}
\int_{-\infty}^{\infty}dy 
\,n\left(\frac{x-q^{z}y}{r^{z}}\right)n(y),
\nonumber
\end{eqnarray}
%\begin{multicols}{2}
\noindent
with
$P\!=\!\lambda^2/(2^d\Gamma(\frac{d-1}{2})\pi^{(d+3)/2})$,
$r^2=1+q^2-2q\cos\theta$.
It will be convenient to write Eqs. (\ref{g},\ref{n})
as a function of time $t$, by Fourier transforming them:
\begin{equation}
{\widehat{\,\,\,\frac{g_R}{\mid 
g\mid^2}\,\,\,}}(t)=\widehat{I_1}(t)
\label{ghat}
\end{equation}
\begin{equation}
\widehat{\,\,\,\frac{n}{\mid g\mid^2}\,\,\,}(t)=\widehat{I_2}(t)
\label{nhat}
\end{equation}
where $\widehat{I_1}$ is the Fourier transform of the real part of 
$I_1$
and $\widehat{I_2}$ is the Fourier transform of $I_2$
%\end{multicols}
\begin{eqnarray}
&
\widehat{I_1}(t)= 2 \pi P\int_{0}^{\pi}d\theta \sin^{d-2}\theta
\int_{0}^{\infty}dq\cos\theta (\cos\theta-q)q^{2z-3}
\widehat{g_R}\left(t r^{z}\right)\widehat{n}(t q^{z}) \,,
\nonumber
\\
&
\widehat{I_2}(t)= \pi P\int_{0}^{\pi}d\theta \sin^{d-2}\theta
\int_{0}^{\infty}dq(\cos\theta-q)^2 q^{2z-3}r^{-(d+4-2z)}
\widehat{n}\left(t r^{z}\right)\widehat{n}(t q^z) \,.
\nonumber
\end{eqnarray}
%%%%%%%%%%%%%%%%%%%%%%%%%%%%%%%%%%%%%%%%%%%%%%%%%%%%%%%%%%%%%%%%%%
%%%%%%%%%%%%%%%%%%%%%%%%%%%%%%%%%%%%%%%%%%%%%%%%%%%%%%%%%%%%%%%%%%
%%%%%%%%%%%%%%%%%%%%%%%%%%%%%%%%%%%%%%%%%%%%%%%%%%%%%%%%%%%%%%%%%%
%%%%%%%%%%%%%%%%%%%%%%%%%%%%%%%%%%%%%%%%%%%%%%%%%%%%%%%%%%%%%%%%%%
%%%%%%%%%%%%%%%%%%%%%%%%%%%%%%%%%%%%%%%%%%%%%%%%%%%%%%%%%%%%%%%%%%
%%%%%%%%%%%%%%%%%%%%%%%%%%%%%%%%%%%%%%%%%%%%%%%%%%%%%%%%%%%%%%%%%%
%\begin{multicols}{2}
Assuming an exponential, or stretched exponential decay for
$\widehat{n}(t)$, we look for an asymptotic solution for
$t\rightarrow \infty$ of Eq. (\ref{nhat}) in the form
\begin{equation}
\widehat{n_\infty}(t)=
A \,(B t)^{\frac{\gamma}{z}}
e^{-\!\mid B t \mid^{\frac{\alpha}{z}}}\,.
\label{largep}
\end{equation}
We will show that such a solution exists and we will determine
the exponents $\gamma=(d-1)/2$ and $\alpha=1$.
Even though the scaling part of the correlation and response 
function
are coupled in the two Eqs. (\ref{ghat}) and (\ref{nhat}),
we will see that the specific form of $\widehat{g}(t)$ is not
relevant to the large $t$ behaviour of $\widehat{n}(t)$.
We will proceed as follows: we evaluate the right hand side
of Eq. (\ref{nhat}) in the large $t$ limit by saddle point 
methods.
The result of this analysis shows that any $\alpha \leq 1$
allows to reproduce the exponential factor.
For $\alpha \leq z$, we can then assume that the left hand side
of Eq. (\ref{nhat}) is dominated by $g(0)^{-2} \widehat{n}(t)$
(see below), thus the asymptotic form of $I_2(t)$ has to be 
matched with
$g(0)^{-2} \widehat{n_\infty}(t)$. The exponential factor
can be matched by any $\alpha \leq 1$, but we will see that
the only value of $\alpha$ that allows also to match the power
law factor $t^{\gamma/z}$  is $\alpha=1$. We can then proceed
with $\alpha =1$ to get $\gamma=\frac{d-1}{2z}$ and
\begin{equation}
A=\frac{g(0)^{-2}4 \Gamma(4z-4)}{P 2^{(d-1)/2}
\Gamma((d-1)/2)\Gamma(2z-2)^2}\,.
\label{A}
\end{equation}

Let us start with the evaluation of the left hand side of
Eq. (\ref{nhat}). Expanding $\mid g(x)\mid^2$ in even powers of
$x$ around $x=0$, we can write
$\widehat{(n/\mid g\mid^2)}(t)=
a_0\widehat{n}(t)+ a_1d^2\widehat{n}(t)/dt^2 +..$ where 
$a_0=g(0)^{-2}$.
For $\widehat{n}(t)$ of the form (\ref{largep}) with $\alpha \leq 
z$,
the first term
of the series will dominate at large $t$, we can therefore
assume that the left hand side of Eq. (\ref{nhat}) has the
asymptotic behaviour $\widehat{(n/\mid g\mid^2)}(t)\simeq 
g(0)^{-2}\widehat{n}(t)$.
Let us now turn to the asymptotic analysis of $\widehat{I_2}(t)$.
First note that the integral $\widehat{I_2}(t)$, is symmetric
in the exchange $q \rightarrow r$, $\theta \rightarrow \phi$, 
where
$r \sin\phi=q \sin\theta $ (see Fig 1). This allows us to rewrite
$\widehat{I_2}(t)$  as twice the integral restricted
to the region $q \cos \theta <1/2$:
%\end{multicols}
\begin{equation}
\widehat{I_2}(t)= 2 \pi P\int_{0}^{\infty}dq
\int_{\bar{\theta}}^{\pi}d\theta \sin^{d-2}\theta
\int_{0}^{\infty}dq(\cos\theta-q)^2 q^{2z-3}r^{-(d+4-2z)}
\widehat{n}\left(t r^{z}\right)\widehat{n}(t q^z) \,,
\label{res}
\end{equation}
%\begin{multicols}{2}
\noindent
where $\cos\bar{\theta}=\max{[1/2q,1]}$.
In $\widehat{I_2}(t)$ the function $\widehat{n}(t)$ appears
in the integrand with the arguments $tq^z$, $tr^z$.
For large enough $t$, we can always safely replace 
$\widehat{n}(tr^z)$
with its asymptotic form $\widehat{n_\infty}(tr^z)$, since in the
new region of integration $r\geq 1/2$.  We have to use more care
with $\widehat{n}(tq^z)$, since such replacement is not allowed
for all $q$, but only for $q > C/t^{1/z}$, where $C$ is a large 
constant.
We then have $\widehat{I_2}(t)\simeq
\widehat{I_2^{\infty}}(t)=J_1(t)+J_2(t)$, where
%\end{multicols}
\begin{eqnarray}
&
J_1(t)= 2 \pi P\int_{0}^{C/t^{1/z}}dq
\int_{\bar{\theta}}^{\pi}d\theta \sin^{d-2}\theta
(\cos\theta-q)^2 q^{2z-3}r^{-(d+4-2z)}
\widehat{n_\infty}\left(t r^{z}\right)\widehat{n}(t q^z) \,,
\nonumber
\\
&
J_2(t)= 2 \pi P\int_{C/t^{1/z}}^{\infty}dq
\int_{\bar{\theta}}^{\pi}d\theta \sin^{d-2}\theta
(\cos\theta-q)^2 q^{2z-3}r^{-(d+4-2z)}
\widehat{n_\infty}\left(t r^{z}\right)\widehat{n_\infty}(t q^z) 
\,.
\nonumber
\end{eqnarray}
\noindent
The contribution from $J_1$ is negligible with respect to $J_2$
as $t\rightarrow \infty$. Let us evaluate an asymptotic expression 
for
$J_2$  and postpone the discussion of $J_1$ to the end.
Inserting $\widehat{n_\infty}(t)=
A \,(B t)^{\gamma/z}
\exp(-\!\mid B t \mid^{\alpha/z})$ in $J_2(t)$ gives
\begin{equation}
J_2(t)= 2 \pi P A^2 
(Bt)^{\frac{2\gamma}{z}}\int_{C/t^{1/z}}^{\infty}dq
\int_{\bar{\theta}}^{\pi}d\theta \sin^{d-2}\theta
(\cos\theta-q)^2 q^{2z-3+\gamma}r^{-(d+4-2z)+\gamma}
e^{-(Bt)^{\frac{\alpha}{z}} (q^{\alpha}+r^{\alpha})} \,.
\nonumber
\end{equation}
It is immediately clear that the main contribution
to this integral will come from the region where
$f(q,\theta)=q^{\alpha}+r(q,\theta)^{\alpha}$ has its minimum,
i.e. from the segment $\theta=0$, $C/t^{1/z}<q<1/2$ (for large
enough $t$, $C/t^{1/z}<1/2$).
The value of $q$ which minimises $f$ will depend on
$\alpha$: for $\alpha < 1$, $f$ reaches its minimum at
$q_0=C/t^{1/z}$, where  $f(q_0,0)=C^{\alpha}/t^{\alpha/z}+ 
\mid(1-C/t^{1/z})\mid^{\alpha}\simeq 1+(C/t^{1/z})^{\alpha} +
O(C/t^{1/z})$.
For $\alpha>1$, the minimum is realized at $q=1/2$, where
$f(1/2,0)=1/2^{\alpha-1}$. For $\alpha =1$, all $q$
in the region $C/t^{1/z}<q<1/2$ contribute equally, giving
$f(q,0)=1$.
The saddle point approximation in the angular integral gives:
\begin{equation}
J_2(t)\simeq 2 \pi P A^2 
(Bt)^{\frac{2\gamma}{z}}\int_{C/t^{1/z}}^{1/2}dq
e^{-(Bt)^{\frac{\alpha}{z}} (q^{\alpha}+\mid 1-q \mid^{\alpha})}
\mid 1-q \mid^{2-(d+4-2z)+\gamma} q^{2z-3+\gamma}
\int_{0}^{\pi}d\theta \theta^{d-2}
e^{-(Bt)^{\frac{\alpha}{z}}\frac{\alpha}{2}q \mid 1-q 
\mid^{\alpha-1}
\theta^2} \,.
\nonumber
\end{equation}
Performing the integral over $\theta$ in the large $t$ limit:
\begin{equation}
J_2(t)\simeq  \pi P A^2 \Gamma \left(\frac{d-1}{2}\right)
\left(\frac{2}{\alpha}\right)^{\frac{d-1}{2}}
(Bt)^{\frac{2\gamma}{z}- \frac{d-1}{2z}}
\int_{C/t^{1/z}}^{1/2}dq
e^{-(Bt)^{\frac{\alpha}{z}} (q^{\alpha}+\mid 1-q \mid^{\alpha})}
\mid 1-q \mid^{\phi}
q^{\psi}
\,,
\nonumber
\end{equation}
where $\phi=2z-d-2+\gamma+(1-\alpha)\frac{d-1}{2}$ and
$\psi=2z-3+\gamma-\frac{d-1}{2}$.
The leading contribution for large $t$ is then given by
\begin{equation}
J_2(t)\simeq
\pi P A^2 \Gamma \left(\frac{d-1}{2}\right)
\left(\frac{2}{\alpha}\right)^{\frac{d-1}{2}}
(Bt)^{\frac{2\gamma}{z}- \frac{d-1}{2z}}
\left\{
\begin{array}{ll}
\frac{\Gamma\left(\frac{\psi+1}{\alpha}\right)}{\alpha}
\,\,
e^{-(Bt)^{\frac{\alpha}{z}}}
(Bt)^
{-\frac{\psi+1}{z}} & \alpha < 1 \\
I
\,\,
e^{-(Bt)^{\frac{1}{z}}} & \alpha=1 \\
\sqrt{\pi}2^{-(\phi+\psi)}
\,\,
e^{-\frac{(Bt)^{\frac{\alpha}{z}}}{2^{\alpha-1}}}
(Bt)^{-\frac{\alpha}{2z}} & \alpha>1
\end{array}
\right.
\end{equation}
where $I=\int_0^{1/2}dq
q^{\psi} (1-q)^{\phi}$.
For $\alpha \leq 1$, the left hand side of
Eq. (\ref{nhat}) will be dominated by a term of the form
$g(0)^{-2}\widehat{n_{\infty}}(t)$, thus $\widehat{I_2(t)}\simeq 
J_2(t)$
must be proportional to $\widehat{n_{\infty}}(t)$.
This can only be achieved with $\alpha=1$ and 
$\gamma=\frac{d-1}{2}$.
For $\alpha<1$, the stretched exponential factor is
recovered, but trying to match the power of $t$ leads to
$\gamma=\frac{d-1}{2}+\psi+1$, which can only be satisfied
by the unphysical value $z=1$.
For $1 <\alpha <z $, the the left hand side of
Eq. (\ref{nhat}) cannot be matched by $J_2$.
We then conclude that $\alpha$ must be equal to $1$, and the
asymptotic solution is then given by Eq. (\ref{largep}).
The coefficient $A$ can be determined as well by observing that
$\gamma=\frac{d-1}{2}$ gives $\psi=\phi=2z-3$, and thus
$I=\Gamma(2z-2)^2/2\Gamma(4z-4)$ which leads to Eq. (\ref{A}).

\begin{multicols}{2}
Let us now go back to the integral $J_1$.
The argument of $\widehat{n}(tq^z)$ in $J_1$ runs between $0$ and 
$C^z$.
Thus we can obtain an upper bound on $J_1$ by replacing
$\widehat{n}(tq^z)$ by its maximum in this region (which
is given by $\widehat{n}(0)$ if $\widehat{n}(t)$ is a
monotonically decreasing function).
An analysis similarly to the one done for $J_2$, can then be
performed for $J_1$, leading to
$J_1 \propto (Bt)^{\frac{\gamma}{z}-\frac{d-1}{2z}}
e^{-(Bt)^{\frac{\alpha}{z}}}$. Thus the contribution to 
$\widehat{I_2^{\infty}}$
from $J_1$ is down by a factor $(Bt)^{-\frac{\gamma}{z}}$.

The scale parameter $B$ is just that, and remains unfixed 
in terms of the parameters of the KPZ equation. In a recent work 
\cite{us}, we proposed a scaling
ansatz for the correlation function as $z\rightarrow 2$.
If such an ansatz is correct, in a parametrization where $g(0)$ 
is finite the scale $B^{-1}$ on which $t$ varies
would go to zero as $z$ approaches $2$: $B^{-1} \simeq (2-z)$.
Thus the asymptotic solution that we have found here
could be read as an asymptotic solution for all $t$
as $z$ approaches $2$.

In summary, we have presented an analytical derivation
of an asymptotic solution of the mode-coupling equations
for the strong coupling regime of the KPZ equation
in the large $t$ limit. We predict a stretched exponential 
relaxation
for the correlation function, with a power law prefactor. We hope 
that this prediction will stimulate  numerical investigations of 
long-time limits generally.

The authors acknowledge  the support of EPSRC under grants 
GR/L38578 and GR/L97698.
%%%%%%%%%%%%%%%%%%%%%%%%%%%%%%%%%%%%%%%%%%%%%%%%%%%%%%%%%%%%%%%%%%
%%%%%%%%%%%%%%%%%%%%%%%%%%%%%%%%%%%%%%%%%%%%%%%%%%%%%%%%%%%%%%%%%%
%%%%%%%%%%%%%%%%%%%%%%%%%%%%%%%%%%%%%%%%%%%%%%%%%%%%%%%%%%%%%%%%%%
%%%%%%%%%%%%%%%%%%%%%%%%%%%%%%%%%%%%%%%%%%%%%%%%%%%%%%%%%%%%%%%%%%
%%%%%%%%%%%%%%%%%%%%%%%%%%%%%%%%%%%%%%%%%%%%%%%%%%%%%%%%%%%%%%%%%%
%%%%%%%%%%%%%%%%%%%%%%%%%%%%%%%%%%%%%%%%%%%%%%%%%%%%%%%%%%%%%%%%%%

\end{multicols}
\begin{figure}
\centerline{
\epsfxsize=3.1in
\epsfbox{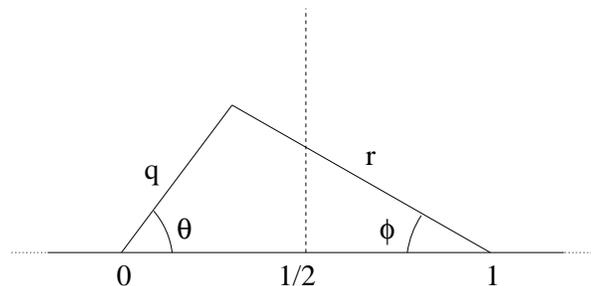}}
\caption{Region of integration for $\widehat{I_2}(t)$ in Eq. (\ref{res}).}
\end{figure}
%\end{multicols}
\end{document}